\documentclass[showpacs,pre,amssymb,amsmath,preprint]{revtex4}

\usepackage{psfrag}
\usepackage{amsmath}
\usepackage{amssymb}
\usepackage{textcomp}		
\usepackage{array}		
\usepackage{multirow}		
\usepackage{rotating}		
\usepackage{floatflt}
\usepackage{textcomp}		
\usepackage{color}

\numberwithin{equation}{section}

\newcommand{\be}{\begin{equation}}
\newcommand{\ee}{\end{equation}}
\newcommand{\bea}{\begin{eqnarray}}
\newcommand{\eea}{\end{eqnarray}}
\def \la{\label}

\def\({\left (}
\def\){\right )}
\def\]{\right]}
\def\[{\left[}
\def\<{\left <}
\def\>{\right>}

\newcommand{\bF}{\mathbf{F}}
\newcommand{\bx}{\mathbf{x}}

\newcommand{\br}{\mathbf{r}}

\newcommand{\bR}{\mathbf{R}}

\renewcommand{\d}{\mathrm{d}}

\newcommand{\na}{\pmb{\nabla}}

\begin{document}

  \title{Algebraic infection of charge correlations of a classical electrolyte at the 
	critical point of the liquid-gas transition}

\author{Angel Alastuey$^{1}$ and Subir K. Das$^2$ 
       \\[2mm]
       $^1${\small Laboratoire de Physique, ENS Lyon, UMR CNRS 5672}\\[-1mm]
        {\small 46 all\'ee d'Italie, 69364 Lyon Cedex 07, France}\\[-1mm]
              $^2${\small Theoretical Sciences Unit and School of Advanced Materials, Jawaharlal Nehru Centre for Advanced Scientific Research, Jakkur P.O., Bangalore 560064,} \\[-1mm] 
							{\small India}}

  \date{\today}
\begin{abstract}
We consider a classical Two-Component Plasma analog of the Restricted Primitive Model of electrolyte, where the hard-core interaction is 
replaced by a soft differentiable potential. Within the Born-Green-Yvon hierarchy for the equilibrium distribution functions, 
we shed light on an infection mechanism where the charge correlations are polluted by the density correlations at the critical point 
of the liquid-gas transition. This implies an algebraic decay of critical charge correlations. Such breakdown of exponential 
clustering should provide dielectric rather than conducting properties at the critical point, leading to the violation of certain charge-charge sum rules. This is in agreement with Monte Carlo simulations. 

\end{abstract}
\maketitle



\section{Introduction}

The liquid-gas transition of a classical electrolyte is a long standing problem, which has been nicely reviewed by 
Fisher~\cite{Fisher}. This has motivated numerous experimental and theoretical works. On the 
theoretical side, there still remain difficult questions which have not been 
satisfactorily answered. A first widely debated problem is about the universality class of the behavior of thermodynamic properties near the critical point \cite{SimulationRPM}. 
It was believed, over many years, that the critical exponents are given by mean-field 
approaches. However, since the end of the last century, careful numerical and experimental inspections at the immediate vicinity of the critical point 
strongly suggest that these exponents are of the Ising-type. There exits now a reasonably general 
consensus that the transition does belong to the Ising universality class \cite{IsingClass}.
 
\bigskip

A second controversial question is by concerning the conducting versus dielectric nature of the critical point, in particular for 
the fully (charge and size) symmetric Restricted Primitive Model (RPM) \cite{Fisher, DKF2012}. Away from the critical point, numerical simulations, 
see \textit{e.g.}~\cite{DKF2012} or~\cite{CaillolLevesque2014}, have
convincingly shown that both the liquid and the gas phases  
display perfect screening properties that are typical of conducting phases. 
However, a first suspicion about the breakdown of perfect screening properties, close to 
the critical point, was pointed out by Caillol~\cite{Caillol1995}. 
More recently, such breakdown was confirmed through sophisticated Monte-Carlo simulations~\cite{DKF2012, DasKimFisher2011}: 
they show via state-of-the-art finite-size scaling analysis \cite{FSS} that the second moment Stillinger-Lovett (SL) sum rule~\cite{StiLov} for equilibrium charge correlations is violated. This 
implies that infinitesimal external charges are no longer perfectly screened (see e.g.~\cite{Martin1988}). Here note that the charge-charge structure factor can be written in even powers of the wave number $k$. The square root of the coefficient of $k^2$, the second moment charge-charge correlation length, equals the Debye screening length \cite{Fisher}. This is a consequence of screening and is referred to as the second SL sum rule. The first SL sum rule dictates that the structure factor is zero at $k=0$. This emerges from the charge neutrality in the system. 

\bigskip

The large-distance decay of charge-charge correlations, $S(r)$, at the critical point is a central question, directly related 
to the system's conducting or dielectric nature. If the density-density correlations decay as a power law~\cite{Fisher}, 
namely as $1/r^{1+\eta}$ with $\eta > 0$, various theoretical approaches predict an exponential decay of equilibrium
charge correlations for the RPM. In particular, this is observed in a solvable mean-spherical model~\cite{AquaFisher2004}, which is expected to 
share common properties with the RPM. In this paper, we consider a fully symmetric Two-Component Plasma (TCP), which is the 
analog of the RPM where the hard core interaction is replaced by a smooth differentiable potential. Our main purpose is to show 
that the critical charge correlations for this TCP are infected by the slow algebraic decay 
of the critical density correlations. In other words, the charge correlations also decay in a power law fashion. 
This is achieved through an inspection of the large-distance behavior of the equilibrium Born-Green-Yvon (BGY) equations. Our analysis sheds light on the 
infection mechanism, alongside providing the power law decay of $S(r)$ as $1/r^{6+\eta}$. Note that within the 
solvable mean-spherical model~\cite{AquaFisher2004}, an infection mechanism occurs for its charge-asymmetric version but not for the  
symmetric version analog to the present model and the RPM.    
An analogous coupling between the charge-charge and density-density correlations can possibly be realized, even for the RPM, within the generalized Debye-H\"uckel theory of Lee and Fisher, via consideration of charge fluctuation that complements the density fluctuations.

\bigskip

In Section~\ref{CSIF}, we describe the considered TCP model.
Like the RPM, this model is expected to undergo a liquid-vapor transition at low density, general features of which
are briefly stated. The BGY hierarchy is introduced in Section~\ref{BGY}. We argue that the corresponding BGY equations should remain valid 
at the critical point. The infection mechanism which results from the coupling between charge and density correlations is highlighted in 
Section~\ref{sec:S3}. We show how this mechanism prevents the exponential decay of $S(r)$. A plausible algebraic-decay 
scenario for two-and three-body particle correlations is shown to be consistent with the BGY equations in Section~\ref{AlgebraicDecay}. 
Accordingly $S(r)$ is found to decay as $1/r^{6+\eta}$. We briefly discuss in Section~\ref{Concluding} how our results 
strongly suggest that the critical point is indeed dielectric, in agreement with numerical simulations~\cite{DasKimFisher2011}.

\section{Charge symmetric ionic fluid}
\la{CSIF}

\subsection{Pairwise interactions}

We consider a two-component classical plasma (TCP) made of two species $(\alpha=+,-)$ of mobile particles, carrying 
charges $\pm q$, in space dimension $d=3$. The particles interact \textsl{via } a sum of pairwise interactions
\begin{align}
\label{Potential}
& u_{++}(r)=u_{--}(r)= v_{\rm SR}(r) + q^2 v_{\rm C}(r)   \nonumber \\
& u_{+-}(r)=u_{-+}(r)= v_{\rm SR}(r) - q^2 v_{\rm C}(r) \; ,  
\end{align}
which include the familiar Coulomb potential $v_{\rm C}(r)=1/r$ and a short-range repulsive interaction $v_{\rm SR}(r)$. 
The short-range interaction diverges positively faster than $1/r$ when $r \to 0$ in order to avoid the collapse between oppositely charged particles.
A possible choice for this is 
\be
\la{PotentialSR}
v_{\rm SR}(r)=V_0 \left[\frac{\sigma}{r}\right]^{12} \exp(-r/\sigma),
\ee
with $V_0 >0$. In the following, the analysis will be performed for general forms of $v_{\rm SR}(r)$ and is not specific to the 
choice~(\ref{PotentialSR}).

\bigskip

We restate, the present fully symmetric TCP is quite similar to the celebrated RPM, 
which is fully symmetric with respect to the charges $\pm q$ and the hard-core diameters $\sigma$. Here the hard-core potential is replaced 
by a soft form which is differentiable everywhere, except at $r=0$. This allows us to introduce the BGY hierarchy
as described in the next Section~\ref{BGY}.  

\subsection{Liquid-vapor phase transition}

In this paper, we assume that the thermodynamic limit (TL) of the present model exists for any choice of $v_{\rm SR}(r)$. Note that
this has been proven only for short-range regularizations of the Coulomb interaction~\cite{FroPar1978} (see also the review~\cite{BryMar1999}). 
Overall charge neutrality in the system is imposed. Not at too low temperatures, the system in a fluid phase 
with a common uniform particle density $\rho$ for both species, is invariant under 
translations once the TL has been taken. As suggested by numerical simulations for similar systems~\cite{SimulationTCP}, like the RPM 
for instance~\cite{SimulationRPM}, the present model is expected to undergo a liquid-vapor phase transition. 
Similar to what occurs for classical fluids made with neutral particles and Lenard-Jones interactions~\cite{SimulationLJ}, 
at a fixed temperature ($T$) lower than some critical value ($T_c$), depending upon the overall density, in equilibrium there will be coexistence of vapor and liquid phases with densities $\rho_{\rm vap}(T)$ and 
$\rho_{\rm liq}(T)$. At the critical point $(T_{\rm c},\rho_{\rm c})$, $\rho_c$ being the critical value of overall density, $\rho_{\rm vap}$ and $\rho_{\rm liq}$ become identical, i.e.,
 $\rho_{\rm c} =\rho_{\rm liq}(T_{\rm c}) = \rho_{\rm vap}(T_{\rm c})$.

\bigskip

In the vapor phase, at low densities, particle correlations should decay exponentially fast at large distances. This is 
strongly suggested by Debye theory, and also by systematic corrections to this mean-field approach derived within the Abe-Meeron 
diagrammatic expansions~\cite{Abe,Meeron} (see also the rigorous proof by Brydges and Federbush~\cite{BryFed1980} for similar systems). 
Perfect screening of external charges 
is then observed, as encoded in the SL second moment sum rule~\cite{StiLov} concerning the charge correlations. When the density increases, the 
system is expected to remain in a conducting state for both vapor and liquid phases, with perfect screening properties. Note that although 
particle correlations then might decay slower than an exponential~\cite{AM1985}, sufficiently fast power-law decays ensure the 
validity of the SL rule~\cite{GruMar1983} (see also the review~\cite{Martin1988}). 
Moreover, the numerical simulations show the persistence of free charges which in turn ensure
that screening properties still hold. However, as shown via sophisticated Monte-Carlo simulations for the RPM~\cite{DasKimFisher2011}, 
this picture becomes quite doubtful 
at the critical point $(T_{\rm c},\rho_{\rm c})$. As a consequence of the coupling between fluctuations in particle and charge correlations, the latter 
is expected to be infected by the slow power-law decay of critical particle correlations, and 
ultimately the second SL sum rule is violated. The main 
purpose of the present paper is to analyze the infection mechanism within the BGY hierarchy.

\section{The BGY hierarchy}
\la{BGY}

The BGY hierarchy should be \textsl{a priori} valid for the distribution functions of any infinitely extended equilibrium state, 
provided that the involved spatial integrals do converge in the infinite space. We first write the second BGY equation for the pair 
distribution functions $\rho_{\alpha_1\alpha_2}(\br_1,\br_2)$ of the infinite system in an homogeneous fluid phase. 
Then we show that all terms are well-behaved if we assume weak clustering properties for the particle correlations, which 
are consistent with their expected large-distance decays in the fluid phase, including the critical point. This confirms that the BGY 
hierarchy can be safely used, even at the critical point.

\subsection{The second BGY equations}

In a fluid phase, the particle distribution functions are invariant under translations, and the second BGY equations 
for the pair distribution functions $\rho_{++}(\br_1,\br_2)=\rho_{--}(\br_1,\br_2)$ and $\rho_{+-}(\br_1,\br_2)=\rho_{-+}(\br_1,\br_2)$, 
can be written by fixing one particle at the origin, $\br_1=\mathbf{0}$, and by taking the gradient with respect to the position $\br_2=\br$ of the 
second particle. This provides
\begin{multline}
\la{2BGYa}
\na \rho_{++}(\mathbf{0},\br) = \beta \rho_{++}(\mathbf{0},\br) \bF_{++}(\mathbf{0},\br) \\
+ \beta \int \d \br' \left[ \rho_{+++}(\mathbf{0},\br,\br') \bF_{++}(\br',\br)+ \rho_{++-}(\mathbf{0},\br,\br') \bF_{-+}(\br',\br) \right] \; , 
\end{multline} 
and
\begin{multline}
\la{2BGYb}
\na \rho_{+-}(\mathbf{0},\br) = \beta \rho_{+-}(\mathbf{0},\br) \bF_{+-}(\mathbf{0},\br) \\
+ \beta \int \d \br' \left[ \rho_{+-+}(\mathbf{0},\br,\br') \bF_{+-}(\br',\br)+ \rho_{+--}(\mathbf{0},\br,\br') \bF_{--}(\br',\br) \right] \; . 
\end{multline}
In these equations, $\bF_{\alpha_1\alpha_2}(\br',\br)=-\na_{\br} u_{\alpha_1\alpha_2}(\br',\br)$ is the force exerted 
on a particle with species $\alpha_2$ and position $\br$, 
by a particle with species $\alpha_1$ and position $\br'$. This force can be decomposed as
\begin{align}
\la{ForceDecomposition}
&\bF_{++}(\br',\br) = \bF_{\rm SR}(\br-\br') + q^2 \bF_{\rm C}(\br-\br') \nonumber \\ 
&\bF_{-+}(\br',\br) = \bF_{\rm SR}(\br-\br') - q^2 \bF_{\rm C}(\br-\br') 
\end{align}
with the short-range part
\be
\la{ForceSR}
\bF_{\rm SR}(\br-\br')= - \na_\br v_{\rm SR} (\br-\br')
\ee
and the Coulomb part
\be
\la{ForceC}
\bF_{\rm C}(\br-\br')= - \na_\br v_{\rm C} (\br-\br')=\frac{\br-\br'}{|\br-\br'|^3} \; .
\ee

\bigskip

For further purposes, it is useful to express the two- and three-body distribution functions in terms of the 
corresponding particle correlations, whose dimensionless counterparts are the Ursell functions, namely
\begin{align}
\la{Ursell}
& \rho_{\alpha_1\alpha_2}(\mathbf{0},\br) = \rho^2 [1 + h_{\alpha_1\alpha_2}(r)] \nonumber \\
& \rho_{\alpha_1\alpha_2\alpha_3}(\mathbf{0},\br,\br') = \rho^3 [1 + h_{\alpha_1\alpha_2}(r) + h_{\alpha_1\alpha_3}(r') + h_{\alpha_2\alpha_3}(|\br'-\br|) 
+ h_{\alpha_1\alpha_2\alpha_3}^{(3)}(\mathbf{0},\br,\br')  ] \; .
\end{align}
Thanks to both the translational and rotational invariance of the fluid phase, the two-body Ursell functions $h_{\alpha_1\alpha_2}$ only depend on the relative 
distance between the fixed particles. Similarly, the three-body Ursell function $h_{\alpha_1\alpha_2\alpha_3}^{(3)}(\mathbf{0},\br,\br') $
only depends on the relative distances $(r,r',|\br'-\br|)$, i.e. the three sides of the triangle formed by the particles.
Using the decompositions~(\ref{ForceDecomposition}) and~(\ref{Ursell}) of the force and of the distribution functions, we recast the 
BGY equations~(\ref{2BGYa},\ref{2BGYb}) as 
\begin{multline}
\la{2BGYaBis}
\na h_{++}(r) = \beta h_{++}(r) \bF_{\rm SR}(\br) + \frac{\beta}{2 \rho} \int \d \br' N(\br') \bF_{\rm SR}(\br-\br') \\
+ \beta \rho \int \d \br' \left[ h_{+++}^{(3)}(\mathbf{0},\br,\br') +  h_{++-}^{(3)}(\mathbf{0},\br,\br')  \right] \bF_{\rm SR}(\br-\br') \\
+  \beta q^2 h_{++}(r) \bF_{\rm C}(\br) + \frac{\beta}{2 \rho} \int \d \br' S(\br') \bF_{\rm C}(\br-\br') \\
+ \beta q^2 \rho \int \d \br' \left[ h_{+++}^{(3)}(\mathbf{0},\br,\br') -  h_{++-}^{(3)}(\mathbf{0},\br,\br')  \right] \bF_{\rm C}(\br-\br') \; , 
\end{multline} 
and
\begin{multline}
\la{2BGYbBis}
\na h_{+-}(r) = \beta h_{+-}(r) \bF_{\rm SR}(\br) + \frac{\beta}{2 \rho} \int \d \br' N(\br') \bF_{\rm SR}(\br-\br') \\
+ \beta \rho \int \d \br' \left[ h_{+-+}^{(3)}(\mathbf{0},\br,\br') +  h_{+--}^{(3)}(\mathbf{0},\br,\br')  \right] \bF_{\rm SR}(\br-\br') \\
-  \beta q^2 h_{+-}(r) \bF_{\rm C}(\br) - \frac{\beta}{2 \rho} \int \d \br' S(\br') \bF_{\rm C}(\br-\br') \\
- \beta q^2 \rho \int \d \br' \left[ h_{+-+}^{(3)}(\mathbf{0},\br,\br') -  h_{+--}^{(3)}(\mathbf{0},\br,\br')  \right] \bF_{\rm C}(\br-\br') \; . 
\end{multline} 
In these equations, $N(\br')$ is the correlation between  particle densities at points $\mathbf{0}$ and $\br'$, 
\be
\la{TotalPPCorrelation}
N(\br')=2 \rho^2 \left[ h_{++}(r') + h_{+-}(r') \right] +2 \rho \delta(\br') \; ,
\ee
while $S(\br')$ is the correlation between charge densities  at points $\mathbf{0}$ and $\br'$, 
\be
\la{TotalCCCorrelation}
S(\br')=2 q^2 \rho^2 \left[ h_{++}(r') - h_{+-}(r') \right] +2 q^2 \rho \delta(\br') \; .
\ee

\bigskip

The BGY equations for the density-density and charge-charge correlations are 
readily obtained by combining equations~(\ref{2BGYaBis}) and~(\ref{2BGYbBis}), and they become for $\br \neq \mathbf{0}$
\begin{multline}
\la{BGYN}
\na N(\br) = \beta N(\br) \bF_{\rm SR}(\br) + 2\beta \rho \int \d \br' N(\br') \bF_{\rm SR}(\br-\br') 
+ 2\beta  \int \d \br' H_{\rm dd}^{(3)}(\mathbf{0},\br,\br')  \bF_{\rm SR}(\br-\br') \\
+  \beta S(\br) \bF_{\rm C}(\br) + 
2 \beta \int \d \br' H_{\rm dc}^{(3)}(\mathbf{0},\br,\br') \bF_{\rm C}(\br-\br') \; , 
\end{multline} 
and
\begin{multline}
\la{BGYS}
\na S(\br) = \beta S(\br) \bF_{\rm SR}(\br) +  2\beta q^2 \int \d \br'  H_{\rm cd}^{(3)}(\mathbf{0},\br,\br')  \bF_{\rm SR}(\br-\br')  \\
+  \beta q^4 N(\br) \bF_{\rm C}(\br) + 2\beta q^2 \rho \int \d \br' S(\br') \bF_{\rm C}(\br-\br') 
+ 2 \beta q^4 \int \d \br' H_{\rm cc}^{(3)}(\mathbf{0},\br,\br') \bF_{\rm C}(\br-\br') \; . 
\end{multline} 
The various three-body correlations $H^{(3)}$  are defined as linear combinations of the three-body Ursell functions,
\begin{align}
\la{3H}
&H_{\rm dd}^{(3)} =  \rho^3 \left[ h_{+++}^{(3)} +  h_{++-}^{(3)} +
h_{+-+}^{(3)} +  h_{+--}^{(3)} \right]  \nonumber \\
&H_{\rm dc}^{(3)} =  \rho^3 \left[ h_{+++}^{(3)} + h_{+--}^{(3)}
-  h_{++-}^{(3)} -h_{+-+}^{(3)} \right] \nonumber \\ 
&H_{\rm cd}^{(3)}= \rho^3 \left[ h_{+++}^{(3)} +  h_{++-}^{(3)}  
-h_{+-+}^{(3)} -  h_{+--}^{(3)} \right] \nonumber \\
&H_{\rm cc}^{(3)}= \rho^3 \left[ h_{+++}^{(3)} + h_{+-+}^{(3)} -  h_{++-}^{(3)} -  h_{+--}^{(3)} \right] \; ,
\end{align}
and they are related to the equilibrium averages of products of three microscopic particle-density or charge-density operators. 
As shown by the structure of equations~(\ref{BGYN},\ref{BGYS}), the density and charge correlations are coupled together, as it 
can be \textsl{a priori} expected.

\subsection{Validity of the BGY equations at the critical point}

The various integrals over $\br'$ involved in the BGY equations~(\ref{BGYN}) and~(\ref{BGYS}) do converge under rather weak clustering assumptions on 
the decay of two- and three-body particle correlations. Indeed, since the short-range force decays as an exponential at 
large distances $|\br'-\br|$, the integrals 
upon $\br'$ of $\bF_{\rm SR}(\br'-\br)$ times particle correlations are always well behaved. This is not the case of the integrals with 
the Coulomb force since $\bF_{\rm C}(\br'-\br)$ decays as $1/|\br'-\br|^2$ when $\br'$ is separated from a fixed $\br$ by infinite distance.
In order to ensure the (absolute) convergence of the related integrals, the correlations $S(\br')$, 
$ H_{\rm dc}^{(3)}(\mathbf{0},\br,\br')$ and $ H_{\rm cc}^{(3)}(\mathbf{0},\br,\br')$ 
have to decay faster than $1/|\br'|^{1+ \epsilon}$ with  $\epsilon >0$ when $|\br'| \to \infty$.
For any equilibrium state where such weak algebraic decays hold, all terms in the BGY equations are finite: this strongly 
suggests that these equations are indeed satisfied by the corresponding equilibrium particle correlations.

\bigskip

At the critical point, one expects a slow algebraic decay of all $n$-body Ursell functions, with $n=2,3,...$,
typically as $1/r^{1+\eta}$ with a strictly positive exponent $\eta>0$. Hence, correlations 
$S(\br')$, $ H_{\rm dc}^{(3)}(\mathbf{0},\br,\br')$ and $ H_{\rm cc}^{(3)}(\mathbf{0},\br,\br')$ decay at least as 
$1/|\br'|^{1+\eta}$ when $|\br'| \to \infty$ since they are linear combinations of two- and three-body Ursell functions. 
According to the previous analysis, this implies that the BGY equations remain valid at the critical point.

\section{The infection mechanism}	\label{sec:S3}

In order to extract constraints from 
the BGY equations for $N(r)$ and $S(r)$, we first introduce weak assumptions for the respective decays of two- and three-body 
Ursell. Such assumptions are shown to be consistent with the internal charge sum rules which are expected to hold in any 
phase and at the critical point (Section~\ref{Clustering}). Then, we highlight an infection mechanism  
which prevents the exponential decay of $S(\br)$ (Section~\ref{BreakdownExponential}).

\subsection{Clustering assumptions and charge sum rules}
\la{Clustering}

At the critical point, according to their respective definitions of particle density~(\ref{TotalPPCorrelation}) and~charge~(\ref{TotalCCCorrelation}) correlations, 
$N(\br)$ and $S(\br)$ decay at least as 
$1/r^{1+\eta}$ at large distances $r$. Such a decay should hold 
for the density correlations $N(\br)$, in agreement with the divergence of its integral over $\br$, implied by 
the compressibility sum rule (note that here the total density is $2\rho$)
\be
\la{Compressibility}
\int \d \br \; N(\br) = 4 \rho^2 k_{\rm B}T \chi_{T}  \; .
\ee    
Indeed, the isothermal compressibility $\chi_{T}= \left[\rho \partial P/\partial \rho \right]^{-1}$,
where $P$ is the pressure of the system, diverges at the critical point. For the charge correlations 
$S(\br)$, one expects a decay faster than $1/r^{1+\eta}$ in order to satisfy the internal perfect screening rule
\be
\la{IPSR}
\int \d \br \; S(\br) = 0  \; ,
\ee 
which requires the integrability of $S(\br)$ over the whole space. Discarding oscillatory behaviors, this 
implies that $S(\br)$ decays at least as $1/r^{3+ \epsilon}$ with $\epsilon >0$ when $r \to \infty$. This 
leads us to infer
\be
\la{DecayUrsellTwo}
h_{++}(r) \sim \frac{A}{r^{1+\eta}} \quad , \quad h_{+-}(r) \sim \frac{A}{r^{1+\eta}} \quad \text{when} \quad r \to \infty \; , 
\ee
where the common amplitude $A$ does not depend on the charges carried by the particles. 
Note that the charge sum rule~(\ref{IPSR}) is crucial for the consistency of the present picture: 
it guarantees a minimal screening of Coulomb interactions which in turn do not affect the leading critical tails.  

\bigskip

The perfect screening rule~(\ref{IPSR}) means that the total charge carried by the polarization cloud surrounding a 
given fixed particle, exactly cancels its charge. If now two particles with charges $e_{\alpha_1}$ 
and $e_{\alpha_2}$ are fixed at positions $\mathbf{0}$ and $\br$, the total charge 
carried by the corresponding polarization cloud should exactly reduce to $-(e_{\alpha_1}+e_{\alpha_2}) $. Hence, the three-body Ursell 
functions are expected to satisfy the sum rules~\cite{Martin1988}
\be
\la{SumRules3BodyPP}
\int \d \br'\rho \left[ h_{+++}^{(3)}(\mathbf{0},\br,\br') -  h_{++-}^{(3)}(\mathbf{0},\br,\br') \right] = -2 h_{++}(\mathbf{0},\br)
\ee
and
\be
\la{SumRules3BodyPM}
\int \d \br'\rho \left[ h_{+-+}^{(3)}(\mathbf{0},\br,\br') -  h_{+--}^{(3)}(\mathbf{0},\br,\br') \right] = 0 \; .
\ee 
These sum rules imply that the three-body Ursell functions are integrable when two particle positions are fixed while the third one 
is sent to infinity. Discarding oscillatory behaviors as for $S(\br)$, this leads to
\be
\la{DecayUrsellThreePP}
h_{+++}^{(3)}(\mathbf{0},\br,\br') \sim h_{++-}^{(3)}(\mathbf{0},\br,\br') \sim \frac{A_{++}(\mathbf{0},\br)}{|\br' - \br/2|^{1+\eta}} 
\quad \text{when} \quad r' \to \infty \;  
\ee  
and 
\be
\la{DecayUrsellThreePM}
h_{+-+}^{(3)}(\mathbf{0},\br,\br') \sim h_{+--}^{(3)}(\mathbf{0},\br,\br') \sim \frac{A_{+-}(\mathbf{0},\br)}{|\br' - \br/2|^{1+\eta}} 
\quad \text{when} \quad r' \to \infty \; ,
\ee
while the differences $[ h_{+++}^{(3)}(\mathbf{0},\br,\br') -  h_{++-}^{(3)}(\mathbf{0},\br,\br')]$ and 
$[ h_{+-+}^{(3)}(\mathbf{0},\br,\br') -  h_{+--}^{(3)}(\mathbf{0},\br,\br')]$ decay as least at $(1/r')^{3+ \epsilon}$. These 
large-distance behaviors follow from the assumption that the leading critical tails are not affected by the charges of particles. This 
assumption has been used for deriving~(\ref{DecayUrsellTwo}) for two-body correlations. More generally, the correlations between a 
given group of $n > 1$ particles with barycenter $\bR$ on the one hand, and a single particle with position $\br'$ on the other hand, do not depend at leading order when $r' \to \infty$ on the charge of this single particle. The corresponding critical tail reduces to $1/|\br'-\bR|^{1+\eta}$ times 
an amplitude which only depends on the relative distances between the $n$ particles in the considered group. In~(\ref{DecayUrsellThreePP}) 
and~(\ref{DecayUrsellThreePM}), $n=2$ and $\bR=\br/2$, while the functions $A_{++}(\mathbf{0},\br)$ 
and $A_{+-}(\mathbf{0},\br)$ 
can be reasonable assumed to be invariant under rotations, namely they only depend on $r$. Moreover, and similar to the two-body Ursell 
functions $h_{++}$ and $h_{+-}$, they should behave as 
\be
\la{DecayAmplitudeTwo}
A_{++}(r) \sim A_{+-}(r) \sim \frac{A^{(2)}}{r^{1+\eta}} \quad \text{when} \quad r \to \infty \; , 
\ee
while the difference $[A_{++}(r) - A_{+-}(r)]$ decays as least at $1/r^{3+ \epsilon}$.

\bigskip

For further purposes, it is interesting to notice that, by symmetry, the behaviors~(\ref{DecayUrsellThreePP}) and~(\ref{DecayUrsellThreePM}) 
are valid for any triangular configuration where one distance is kept fixed while the two remaining ones diverge, namely
\begin{align}
\la{DecayUrsellThreePPbis}
& h_{+++}^{(3)}(\mathbf{0},\br,\br') \sim \frac{A_{++}(\mathbf{0},\br')}{|\br - \br'/2|^{1+\eta}} \quad \text{when} \quad r \to \infty 
\; , \; \br' \quad \text{fixed}
\nonumber \\
& h_{++-}^{(3)}(\mathbf{0},\br,\br') \sim \frac{A_{+-}(\mathbf{0},\br')}{|\br - \br'/2|^{1+\eta}} \quad \text{when} \quad r \to \infty 
\; , \; \br' \quad \text{fixed}
\nonumber \\
& h_{+++}^{(3)}(\mathbf{0},\br,\br') \sim \frac{A_{++}(\br,\br')}{|(\br + \br')/2|^{1+\eta}} \quad \text{when} \quad r \to \infty
\; , \; (\br'- \br) \quad \text{fixed} 
\nonumber \\
& h_{++-}^{(3)}(\mathbf{0},\br,\br') \sim \frac{A_{+-}(\br,\br')}{|(\br + \br')/2|^{1+\eta}} \quad \text{when} \quad r \to \infty
\; , \; (\br'- \br) \quad \text{fixed} \; ,   
\end{align}  
and 
\begin{align}
\la{DecayUrsellThreePMbis}
& h_{+-+}^{(3)}(\mathbf{0},\br,\br') \sim \frac{A_{++}(\mathbf{0},\br')}{|\br - \br'/2|^{1+\eta}} \quad \text{when} \quad r \to \infty 
\; , \; \br' \quad \text{fixed}
\nonumber \\
& h_{+--}^{(3)}(\mathbf{0},\br,\br') \sim \frac{A_{+-}(\mathbf{0},\br')}{|\br - \br'/2|^{1+\eta}} \quad \text{when} \quad r \to \infty 
\; , \; \br' \quad \text{fixed}
\nonumber \\
& h_{+-+}^{(3)}(\mathbf{0},\br,\br') \sim \frac{A_{-+}(\br,\br')}{|(\br + \br')/2|^{1+\eta}} \quad \text{when} \quad r \to \infty 
\; , \; (\br'- \br) \quad \text{fixed}
\nonumber \\ 
& h_{+--}^{(3)}(\mathbf{0},\br,\br') \sim \frac{A_{--}(\br,\br')}{|(\br + \br')/2|^{1+\eta}} \quad \text{when} \quad r \to \infty 
\; , \; (\br'- \br) \quad \text{fixed} \; .
\end{align} 
Moreover, since the present TCP is fully symmetric, the amplitude functions satisfy the symmetry relations 
$A_{++}=A_{--}$ and $A_{+-}=A_{-+}$.

\bigskip

The sum rule~(\ref{SumRules3BodyPP}) implies another sum rule for the amplitude functions $A_{++}$ and $A_{+-}$.
Let us consider the limit $r \to \infty$ of both sides of~(\ref{SumRules3BodyPP}). In the 
integral in the l.h.s., the leading contributions arise from the regions $\br'$ close to the origin on the
one hand, and $\br'$ close to $\br$ on the other hand. According to the decays~(\ref{DecayUrsellThreePPbis}), 
both regions give identical contributions which lead to 
\be
\la{Integral3bodyPP}
\int \d \br'\rho \left[ h_{+++}^{(3)}(\mathbf{0},\br,\br') -  h_{++-}^{(3)}(\mathbf{0},\br,\br') \right] \sim 2 \rho \; 
\frac{\int \d \bx [A_{++}(x)-A_{+-}(x)]}{r^{1+\eta}} \quad , \quad r \to \infty \; .
\ee
Notice that $[A_{++}(x)-A_{+-}(x)]$ decays as $1/x^{3+ \epsilon}$ like $S(x)$, so 
$\int \d \bx [A_{++}(x)-A_{+-}(x)]$ does converge. Comparing the behavior~(\ref{Integral3bodyPP}) with the 
large-$r$ decay of the r.h.s of~(\ref{SumRules3BodyPP}) inferred from~(\ref{DecayUrsellTwo}), we find 
\be
\la{SumRuleAmplitude}
\rho \int \d \bx [A_{++}(x)-A_{+-}(x)] = -A \; . 
\ee
Analogous manipulations can be repeated for the sum rule~(\ref{SumRules3BodyPM}). The leading contributions of 
regions $\br'$ close to the origin, and $\br'$ close to $\br$, then exactly cancel out by virtue of the 
symmetry relations $A_{++}=A_{--}$ and $A_{+-}=A_{-+}$. Hence, no additional constraints on 
$A_{++}$ and $A_{+-}$ are imposed by sum rule~(\ref{SumRules3BodyPM}). 

\bigskip

We stress that the previous clustering assumptions on the three-body Ursell functions turn out to be perfectly 
consistent with the three-body charge sum rules. Furthermore, they are satisfied by the
Kirkwood superposition approximation~\cite{Kirkwood}, 
\be
\la{WK}
\rho_{\alpha_1\alpha_2 \alpha_3}(\br_1,\br_2,\br_3)= 
\rho_{\alpha_1\alpha_2}(\br_1,\br_2)\rho_{\alpha_1\alpha_3}(\br_1,\br_3)\rho_{\alpha_2\alpha_3}(\br_2,\br_3) \; .
\ee
which provides the three-body Kirkwood Ursell functions 
\be
\la{WK3U}
h_{\alpha_1\alpha_2 \alpha_3}^{(3,{\rm K})}= h_{\alpha_1\alpha_2}h_{\alpha_1\alpha_3} + 
h_{\alpha_1\alpha_2}h_{\alpha_2\alpha_3}
+h_{\alpha_1\alpha_3}h_{\alpha_2\alpha_3} 
+h_{\alpha_1\alpha_2}h_{\alpha_1\alpha_3}h_{\alpha_2\alpha_3} \; .
\ee

\subsection{Breakdown of the exponential decay of charge correlations}
\la{BreakdownExponential}

Within the previous clustering assumptions, we have seen that some combinations of correlations decay faster than $1/r^{1 + \eta}$ because of
cancellations. In particular such mechanism arises for the charge correlations $S(r)$ which should decay at least as  $1/r^{3+ \epsilon}$. 
Let us assume \textit{a priori} that $S(r)$ decays exponentially fast at the critical point. Consistently, we then also assume that
the combinations of three-body correlations, similar to 
$[h_{++}-h_{+-}]$ in $S(r)$, where cancellations of the critical $1/r^{1+\eta}$-tails occur, also decay exponentially fast. 
Using the behaviors~(\ref{DecayUrsellThreePPbis}) and (\ref{DecayUrsellThreePMbis}) for the three-body Ursell functions, 
the corresponding exponential-decay scenario (\textbf{EDS}) reads
\begin{itemize}
\item \textbf{EDS1}: $S(r)$ decays exponentially fast when $r \to \infty$
\item \textbf{EDS2}: $H_{\rm cd}^{(3)}(\mathbf{0},\br,\br')$ decays exponentially fast when $r \to \infty$ with either $\br'$ fixed or  
$(\br'-\br)$ fixed
\item \textbf{EDS3}: $H_{\rm cc}^{(3)}(\mathbf{0},\br,\br')$ decays exponentially fast when $r \to \infty$ with $(\br'-\br)$ fixed
\item \textbf{EDS4}: $[A_{++}(x)-A_{+-}(x)]$ decays exponentially fast when $x \to \infty$
\end{itemize}
Then, the strategy consists in showing that this exponential-decay scenario is not consistent with the large-$r$ behavior of the BGY equation~(\ref{BGYS}).

\bigskip

Let us analyze, within \textbf{EDS}, the large-distance behavior of the various terms in~(\ref{BGYS}). In
the l.h.s. $\na S(\br)$ decays exponentially fast by virtue of \textbf{EDS1}. In the r.h.s., we first consider the two terms involving the short-range force 
$\bF_{\rm SR}$. The direct short-range term 
\be
\la{DirectSR}
\bR_{\rm SR}(\br)=\beta S(\br) \bF_{\rm SR}(\br) 
\ee
obviously decays exponentially fast. Because of the exponential decay of $ \bF_{\rm SR}$, the sole contributions in the three-body short-range term
\be
\la{ThreeBodySR}
\bR_{\rm SR}^{(3)}(\br)=2\beta q^2 \int \d \br'  H_{\rm cd}^{(3)}(\mathbf{0},\br,\br')  \bF_{\rm SR}(\br-\br') 
\ee
which might decay slower than an exponential arising from the region where $\br'$ is close to $\br$. However, because of \textbf{EDS2},  
$H_{\rm cd}^{(3)}(\mathbf{0},\br,\br')$ decays exponentially fast for such configurations. Hence, 
$\bR_{\rm SR}^{(3)}(\br)$ also decays exponentially fast. 

\bigskip

In a second step, we study the three terms which involve the Coulomb force $\bF_{\rm C}$.   
The mean-field term 
\be
\la{MeanField}
\bR_{\rm MF}(\br) = 2\beta q^2 \rho \int \d \br' S(\br') \bF_{\rm C}(\br-\br')
\ee
decays exponentially fast, by virtue of the charge sum rule~(\ref{IPSR}) and of the rotational invariance of $S(\br)=S(r)$. 
The direct Coulomb term 
\be
\la{DirectSR}
\bR_{\rm C}(\br)=\beta q^4 N(\br) \bF_{\rm C}(\br) 
\ee
decays algebraically, namely 
\be
\la{DirectSRbis}
\bR_{\rm C}(\br) \sim 4 \beta q^4 \rho^2 \frac{A}{r^{3+\eta}} \hat{\br} \quad , \quad r \to \infty 
\ee
with $\hat{\br}=\br/r$, discarding exponentially fast decaying corrections. In the three-body Coulomb term 
\be
\la{ThreeBodyC}
\bR_{\rm C}^{(3)}(\br)=2 \beta q^4 \int \d \br' H_{\rm cc}^{(3)}(\mathbf{0},\br,\br') \bF_{\rm C}(\br-\br') \; ,
\ee
there are exponentially decaying contributions from the region $\br'$ close to $\br$ as a consequence of \textbf{EDS3}.
However, there are algebraic contributions from the region $\br'$ close to the origin $\mathbf{0}$ which arise from  
the large-distance behavior 
\be
\la{ThreeBodyCCalgebraic}
H_{\rm cc}^{(3)}(\mathbf{0},\br,\br') \sim  2 \frac{[A_{++}(\mathbf{0},\br')-A_{+-}(\mathbf{0},\br')]}{|\br - \br'/2|^{1+\eta}}  \quad \text{when} 
\quad  r \to \infty \quad \text{with} \quad  \br'  \quad \text{fixed} \; , 
\ee
discarding exponentially decaying terms. Hence, we find 
\be
\la{ThreeBodyCbis}
\bR_{\rm C}^{(3)}(\br)\sim 4 \beta q^4 \rho^3 \int \d \br' \frac{[A_{++}(\mathbf{0},\br')-A_{+-}(\mathbf{0},\br')]}{|\br-\br'/2|^{1+\eta}} \bF_{\rm C}(\br-\br') \; ,
\ee
discarding exponentially decaying terms.

\bigskip

The previous analysis shows that all terms in the BGY equation~(\ref{BGYS}) decay exponentially fast, except the sum
$[\bR_{\rm C}(\br) + \bR_{\rm C}^{(3)}(\br)]$ which, according to~(\ref{DirectSRbis}) and~(\ref{ThreeBodyCbis}) provides the 
algebraic contribution
\be
\la{BGYalgebraic}
 4 \beta q^4 \rho^2 \int \d \br' \frac{[\rho A_{++}(\mathbf{0},\br')-\rho A_{+-}(\mathbf{0},\br') + A \delta(\br')]}{|\br-\br'/2|^{1+\eta}} \bF_{\rm C}(\br-\br') \; .
\ee
At large distances $r$, its asymptotic representation in power series of $1/r$ is generated by the expansion of $\bF_{\rm C}(\br-\br')/|\br-\br'/2|^{1+\eta}$ 
in Taylor series with respect to $\br'$. The \textit{a priori} leading term of order $1/r^{3+\eta}$ has a vanishing amplitude by 
virtue of the sum rule~(\ref{SumRuleAmplitude}). The amplitude of the next term of order $1/r^{4+\eta}$ is proportional to the first moment of 
$[\rho A_{++}(\mathbf{0},\br')-\rho A_{+-}(\mathbf{0},\br') + A \delta(\br')]$ and it also vanishes
because of rotational invariance. The first \textit{a priori} non-vanishing term is of order $1/r^{5+\eta}$ and its amplitude is proportional to the 
second moment of $[\rho A_{++}(\mathbf{0},\br')-\rho A_{+-}(\mathbf{0},\br') + A \delta(\br')]$. Hence the sum
$[\bR_{\rm C}(\br) + \bR_{\rm C}^{(3)}(\br)]$ decays algebraically, namely
\be
\la{BGYalgebraicBis}
 \bR_{\rm C}(\br) + \bR_{\rm C}^{(3)}(\br) = \frac{(1+\eta)(8+\eta) \beta q^4 \rho^2 M_2}{6 \; r^{5+\eta}} \; \hat{\br}  + o(1/r^{5+\eta}) \quad ,
\quad r \to \infty \; 
\ee
with the second moment
\be
\la{SecondMomentAmplitude}
M_2=\rho \int \d \bx \; x^2 \; [A_{++}(x)-A_{+-}(x)] \; .
\ee
If $M_2$ vanishes, one has to pursue the asymptotic large-$r$ expansion of~(\ref{BGYalgebraic}) to next orders. The amplitude of the term of order 
$1/r^{2n+\eta}$ with $n \geq 2$ is proportional to the $2n$-th moment $M_{2n}$ of $[A_{++}(x)-A_{+-}(x)]$. The leading behavior is obtained for the first 
non-vanishing moment $M_{2n}$. Such moment necessarily exists since otherwise $\rho[A_{++}(x)-A_{+-}(x)]$ would reduce to $-A \delta (\bx)$, in contradiction 
with the physical expectation that $\rho[A_{++}(x)-A_{+-}(x)]$ is a smooth function of $x$. Hence, in any case, $[\bR_{\rm C}(\br) + \bR_{\rm C}^{(3)}(\br)]$ 
indeed decays algebraically. 

\bigskip

Since all the other terms than $[\bR_{\rm C}(\br) + \bR_{\rm C}^{(3)}(\br)]$ in~(\ref{BGYS}) decay exponentially fast, we conclude that the assumed 
\textbf{EDS } is not consistent with the BGY hierarchy. Thus, at the critical point, the charge correlations cannot decay exponentially fast, 
and they are polluted by the algebraic tails in the density-density correlations. The infection mechanism arises from the contributions of three-body 
correlations, where the critical $1/r^{1+\eta}$-tails are coupled to the Coulomb $1/r^2$-force: the resulting effective 
$1/r^{3+\eta}$-force created by a spherically symmetric cloud decays algebraically at large distances as far as this cloud displays a finite 
spatial extension.

\section{A plausible scenario}

The infection mechanism described in the previous Section leads to an algebraic decay of charge correlations. In 
Section~\ref{AlgebraicDecay}, we propose an algebraic-decay scenario (\textbf{ADS}), which is consistent with the 
BGY hierarchy, contrary to the \textbf{EDS}. A few concluding comments are given in Section~\ref{Concluding}.

\subsection{Power-law decay of charge correlations}
\la{AlgebraicDecay}

Now we assume that $S(r)$ decays as $1/r^s$ with $s > 3$. Similarly to the \textbf{EDS}, we also assume that the same power-law 
$1/r^s$ controls the decay of the combinations of three-body correlations for configurations where the $1/r^{1+\eta}$-critical tails cancel out. 
Accordingly, the corresponding algebraic-decay scenario reads
\begin{itemize}
\item \textbf{ADS1}: $S(r)$ decays as $1/r^s$ when $r \to \infty$
\item \textbf{ADS2}: $H_{\rm cd}^{(3)}(\mathbf{0},\br,\br')$ behaves as 
\be
\la{ThreeBodyCDalgebraic1}
H_{\rm cd}^{(3)}(\mathbf{0},\br,\br') \sim   \frac{[B_{++}(\mathbf{0},\br')+ B_{+-}(\mathbf{0},\br')]}{|\br - \br'/2|^s}  \quad \text{when} 
\quad  r \to \infty \quad \text{with} \quad  \br'  \quad \text{fixed} \; , 
\ee
\be
\la{ThreeBodyCDalgebraic2}
H_{\rm cd}^{(3)}(\mathbf{0},\br,\br') \sim   \frac{[B_{++}(\br,\br')+ B_{+-}(\br,\br')]}{|(\br + \br')/2|^{s}}  \quad \text{when} 
\quad  r \to \infty \quad \text{with} \quad  (\br'- \br) \quad \text{fixed} \; . 
\ee
\item \textbf{ADS3}: $H_{\rm cc}^{(3)}(\mathbf{0},\br,\br')$ behaves as
\be
\la{ThreeBodyCCalgebraic}
H_{\rm cc}^{(3)}(\mathbf{0},\br,\br') \sim   \frac{[B_{++}(\br,\br')- B_{+-}(\br,\br')]}{|(\br + \br')/2|^{s}}  \quad \text{when} 
\quad  r \to \infty \quad \text{with} \quad  (\br'- \br) \quad \text{fixed} \; . 
\ee 
\item \textbf{ADS4}: $[A_{++}(x)-A_{+-}(x)]$ and $[B_{++}(x)- B_{+-}(x))]$ decay as $1/x^s$ when $x \to \infty$
\end{itemize}
We show that such algebraic decays are consistent with the BGY equations. Furthermore the corresponding 
analysis provides $s=6+\eta$. 

\bigskip

Let us consider the BGY equation~(\ref{BGYS}) for $S(r)$. In
the l.h.s. $\na S(\br)$ decays exponentially as $1/r^{s+1}$ by virtue of \textbf{EDS1}. In the r.h.s., the direct short-range term 
$\bR_{\rm SR}(\br)$ decays exponentially fast because of the exponential decay of $ \bF_{\rm SR}$. In the three-body short-range term
$\bR_{\rm SR}^{(3)}(\br)$, the contributions of the region $\br'$ close to the origin also decay exponentially fast for the same reason. 
According to~(\ref{ThreeBodyCDalgebraic2}), algebraically decaying contributions arise from the region where $\br'$ is close to $\br$. 
The corresponding power series of $1/r$ are 
generated by the expansion of $\bF_{\rm SR}(\br-\br')/|(\br + \br')/2|^{s}$ in Taylor series with respect to $\br'$. 
Combining the rotational invariance of the amplitude functions, $B_{++}(\br,\br')=B_{++}(|\br-\br'|)$ and $B_{+-}(\br,\br')=B_{+-}(|\br-\br'|)$, with
the antisymmetry of $ \bF_{\rm SR}$, we see that the first non-vanishing term decays as $1/r^{s+1}$, and so does $\bR_{\rm SR}^{(3)}(\br)$.

\bigskip

Since 
\be
\la{MeanFieldPoisson}
\na \cdot \bR_{\rm MF}(\br) = - 8 \pi \beta q^2 S(r)
\ee
as a consequence of Poisson's equation, the mean-field term $\bR_{\rm MF}(\br)$ decays as $1/r^{s-1}$. The 
direct Coulomb term $\bR_{\rm C}(\br)$ again decays as~(\ref{DirectSRbis}) discarding terms of order $1/r^{s+2}$. 
In the three-body Coulomb term $\bR_{\rm C}^{(3)}(\br)$, the contribution of the region $\br'$ close to the origin 
again behaves as~(\ref{ThreeBodyCbis}) discarding terms of order $1/r^{s+2}$. The contribution of 
the region $\br'$ close to $\br$ can be determined as that arising in $\bR_{\rm SR}^{(3)}(\br)$ 
by using~(\ref{ThreeBodyCCalgebraic}) and is found to decay as $1/r^{s+1}$. 

\bigskip

In conclusion, the two slowest-decaying contributions in~(\ref{BGYS}) arise from the mean-field term 
$\bR_{\rm MF}(\br)~\sim {\rm cst}/r^{s-1}$ on the one hand, and from the combination 
$[\bR_{\rm C}(\br) + \bR_{\rm C}^{(3)}(\br)] \sim {\rm cst}/r^{5+\eta}$ on the other hand. Thus, 
this BGY equation for $S(r)$ can be satisfied if and only if the two previous powers are identical, 
\textit{i.e.} $s=6+\eta$.

\bigskip

It can be checked that \textbf{ADS} is also consistent with the large-distance behavior of the BGY equation~(\ref{BGYN}) for $N(r)$. 
Indeed, the l.h.s. of~(\ref{BGYN}) decays as $1/r^{2+\eta}$, and in the r.h.s. there are various terms 
which also decay as $1/r^{2+\eta}$, namely the mean-field term and the two three-body terms. Note that the corresponding
$1/r^{2+\eta}$-decays are obtained by combining the antisymmetry of the forces $\bF_{\rm SR}$ and $\bF_{\rm C}$ 
with the asymptotic behavior of $N(r')$, $H_{\rm dd}^{(3)}(\mathbf{0},\br,\br')$ and $H_{\rm dc}^{(3)}(\mathbf{0},\br,\br')$ 
when $r \to \infty$ with $(\br'- \br)$ fixed. Thus, the algebraic-decay scenario (\textbf{ADS} ) is fully consistent with the 
large-distance behavior of the BGY hierarchy, provided that $s=6+\eta$.

\subsection{Concluding comments}
\la{Concluding}

The key ingredients of our derivations are plausible \textit{a priori} assumptions on the decay of 
three-body Ursell functions. In fact, such three-body correlations can be represented by diagrammatic series where the 
bonds are the two-body Ursell functions (see \textit{e.g.}~\cite{DiagrammaticThreeBody}). The algebraic decays of the $h^{(3)}$'s are then related 
to those of the $h$'s. This leads to replace the crude assumptions~(\ref{DecayUrsellThreePP}) and~(\ref{DecayUrsellThreePM}), 
by  asymptotic expansions in inverse power of $r$. However, their structures are identical to those derived from~(\ref{DecayUrsellThreePP}) 
and~(\ref{DecayUrsellThreePM}). Hence, the infection mechanism 
found in Section~\ref{BreakdownExponential} still holds, with a three-body contribution in~(\ref{BGYS}) which again 
decays as $1/r^{5+\eta}$. Including the refined three-body decays in the analysis of Section~\ref{AlgebraicDecay}, 
one finds that $S(r)$ indeed decays as $1/r^{6+\eta}$ when $r \to \infty$. Furthermore, it turns out that such decay 
has been independently obtained in~\cite{AquaFisher} within a completely different approach. Accordingly, the $1/r^{6+\eta}$-decay 
of $S(r)$ appears as a quite robust prediction, despite it is not yet rigorously established. 

\bigskip  

Since $0 <\eta < 1$, $S(r)$ decays faster than $1/r^6$ and slower than $1/r^8$. Hence the second moment of $S(r)$ is finite, while its fourth moment diverges. 
This is in agreement with the numerical results obtained within sophisticated Monte Carlo simulations~\cite{DasKimFisher2011}.
These simulations also indicate that the Stillinger-Lovett second moment sum rule for $S(r)$ is violated at the critical point.
In fact, the slow decays of three- and four-body correlations present in the algebraic-decay scenario should 
prevent the Stillinger-Lovett sum rule to be 
satisfied, as strongly suggested by a conditional theorem~\cite{GruMar1983}. The present analysis turns then to be also consistent 
with the simulation prediction concerning the dielectric nature of the critical point.

\end{document}